\documentclass[twocolumn,eqsecnum,floatfix,aps,prc]{revtex4}
\usepackage{epsfig}
\usepackage{times}
\usepackage{color}

\def\lap{\mathrel{\mathpalette\fun <}}

\def\fun#1#2{\lower3.6pt\vbox{\baselineskip0pt\lineskip.9pt
  \ialign{$\mathsurround=0pt#1\hfil##\hfil$\crcr#2\crcr\sim\crcr}}}
\begin{document}
                    

\title {Universality in Efimov associated tetramers in $^4$H\lowercase{e} 
}

\author{E.\ Hiyama}
\email{hiyama@riken.jp}
\affiliation{RIKEN Nishina Center, RIKEN, Wako 351-0198, Japan}

\author{M.\ Kamimura}
\email{mkamimura@riken.jp}
\affiliation{Department of Physics, Kyushu University,
Fukuoka 812-8581, Japan, \\
RIKEN Nishina Center, RIKEN, Wako 351-0198, Japan}

\date{\today}

\begin{abstract}
We calculated, using
seven~{\it realistic}~$^4$He$\,$-$^4$He potentials
in the literature,
the \mbox{Efimov} spectra of  $^4$He trimer and tetramer and 
analyzed the universality of the systems.
The three-(four-)body Schr\"{o}dinger equations were solved  fully 
nonadiabatically with the high-precision calculation method
employed in our previous work on the $^4$He trimer 
and tetramer [Phys.~Rev.~A~{\bf 85},~022502~(2012);  
{\bf 85},~062505~(2012)].
We found the following universality in the four-boson system: 
i) The critical scattering lengths at which the tetramer ground and 
excited states couple to the four-body threshold
are independent of the choice of the two-body realistic potentials 
in spite of the difference in the short-range details and  
do not contradict the corresponding values
observed in the experiments in ultracold alkali atoms
when scaled with the van der Waals length $r_{\rm vdW}$, and
ii) the four-body hyperradial potential has a repulsive barrier 
at the four-body hyperradius \mbox{$R_4 \approx 3\, r_{\rm vdW}$,} 
which prevents the four particles from  
getting close together to explore nonuniversal features of 
the interactions at short distances.
This result is an extension of the universality in
Efimov trimers that the appearance of the repulsive barrier at 
the three-body hyperradius $R_3 \approx 2\, r_{\rm vdW}$
makes the critical scattering lengths independent of the
short-range details of the interactions as reported in the literature
and also in the present work for the $^4$He trimer with the
realistic potentials. 
\end{abstract}

\pacs{31.15.ac,31.15.xt,67.85.-d}
   
\maketitle

\section{INTRODUCTION}

The universal physics of few particles interacting with
resonant short-range interactions,  
commonly referred to as Efimov physics~\cite{Efimov70},
has intensively studied in recent years both experimentally 
and theoretically (e.g.\cite{Ferlaino-FB11,Braaten06,Braaten07}
for a review). 
The resonant short-range two-body interaction with the
$s$-wave scattering length $a$ at the unitary limit $a \to \pm \infty$
generates the effective three-body attraction that supports
an infinite number of weakly  bound three-body states,
known as the Efimov trimers,
with the peculiar geometric universal scaling of their energies.
For a potential with a finite scattering length $a$,
only a finite number of the three-body bound states exist.
The critical scattering length, $a_-^{(0)}$,
for the appearance of the first Efimov state 
at the three-body threshold on the $a<0$ side, often referred to as 
the three-body parameter, was initially 
considered as a nonuniversal quantity to be affected by the short-range 
details of the interactions. 
The absolute position of $a_-^{(0)}$, which determines
the overall scale of the whole Efimov spectrum, is not predicted 
by the low-energy effective theories~\cite{Efimov70,Braaten06,Braaten07}.

However, several recent  experiments with identical ultracold 
alkali atoms 
suggested that  the three-body parameter $a_-^{(0)}$ might be 
universal since the observed values in 
Refs.~\cite{Kraemer06,Pollack09,Gross09,Wenz09,Ferlaino09,Williums09,
Gross10,Berninger11,Wild12,Dyke13,Zenesini13,Huang14} 
are approximately
the same  in units of the van der Waals length $r_{\rm vdW}$
for different atoms, \mbox{$a_-^{(0)}/r_{\rm vdW} \approx -9.5\pm 15 \%$}  
as summarized in Refs.~\cite{Berninger11,Machtey12,Wang12,Roy13}, 
where $r_{\rm vdW}=\frac{1}{2}(mC_6/\hbar^2)^{1/4}$  
with the atomic mass $m$ and  the 
coefficient $C_6$ of the long-range potential $r^{-6}$~\cite{Chin10}.

This interesting result has stimulated the theoretical studies 
in Refs.~\cite{Wang12,Chin11,Sorensen12,Schmidt12,Naidon12,
Endo12,Tobias2013,Naidon14}. 
In Ref.~\cite{Wang12}, with a numerical study in the
adiabatic  hyperspherical representation,
the universality of the 
three-body parameter $a_-^{(0)}/r_{\rm vdW}$ has been
understood as follows: 
The origin of the universality 
is related to the suppression of the 
probability to find two particles at distances $r \lap r_{\rm vdW}$
where the pairwise interaction features a deep well 
supporting many two-body
bound states or a short-range hardcore repulsion.
This suppression leads to the formation of the universal three-body 
potential barrier around the three-body hyperradius 
$R_3 \approx 2\,r_{\rm vdW}$
(see Eq.~(2.2) below for the definition of the $A$-body hyperradius).
This barrier prevents
the three particles from simultaneously getting close together 
to explore nonuniversal features of 
the interactions at short distances.
Reference~\cite{Endo12} elaborated
on this physical  mechanism by elucidating how the two-body 
suppression leads to appearance of the universal barrier at
$R_3 \approx 2\,r_{\rm vdW}$.

As for the $^4$He trimer interacting 
with a realistic $^4$He potential that has a strong
repulsive core at $r \approx r_{\rm vdW}$ 
and supports only one two-body bound state,
one of the present authors (E.H.) and the collaborators~\cite{Naidon12}
showed, by performing a nonadiabatic three-body calculation, 
$a_-^{(0)}$ is consistent with 
the observed value of the three-body parameter in alkali atoms
when scaled with the effective range.

One of the aims of the present paper is,
using the same framework, 
the Gaussian expansion method 
(GEM)~\cite {Kamimura88, Kameyama89, Hiyama03, Hiyama2012-FBS,
Hiyama2012-PTEP} for few-body systems, 
as in our previous papers~\cite{Hiyama2012} and \cite{Hiyama2012QED}
(hereafter referred to as paper I and \mbox{paper II,} respectively),
to calculate the Efimov spectra of the
$^4$He trimer with the seven different realistic 
$^4$He potentials~\cite{LM2M2,TTY,HFD-B,B3FCI1a,B3FCI1b,SAPT2,
SAPT96a,SAPT96b,CCSAPT07,PCKLJS,PCKLJS-2}. 
We show that the calculated three-body parameters
$a_-^{(v)}$ 
for the ground $(v=0)$ and  excited $(v=1)$  states
are independent of the difference in the short-range details
of the realistic potentials
and are consistent  with the observed values 
of the first and second Efimov resonances in the experiments for
the alkali trimers when scaled with the van der Waals length.  

From the $^4$He trimer wave functions
described with the three sets of the Jacobi coordinates,
we derive the three-body potential $U_3(R_3)$ 
as a function of the three-body hyperradius $R_3$ and show that 
$U_3(R_3)$ has a repulsive core  at  
$R_3 \approx 2 \, r_{\rm vdW}$ and the three-body probability density
inside the core is heavily suppressed
both in the ground and excited states.
This supports, from the view point of the realistic interactions,
the finding~\cite{Wang12,Endo12} of the
appearance of the three-body repulsion universally at
$R_3 \approx 2 \, r_{\rm vdW}$ in the first Efimov trimer state.

As long as the tetramers of the alkali atoms are concerned, 
recent experiments have observed the scattering lengths
$a_-^{(4,0)}$ and $a_-^{(4,1)}$,
at which the first and second universal four-body states tied to 
the first Efimov trimer state emerges at the four-body threshold
on the $a<0$ side. The data~\cite{Ferlaino09,Ferlaino-FB11,
Zenesini13,Dyke13} exhibit a universal property
$a_-^{(4,0)}/r_{\rm vdW} \approx -(3 - 4.5)$ and 
$a_-^{(4,1)}/r_{\rm vdW} \approx - (7 - 8.5)$. 

Another aim of this paper is 
to  calculate the four-body Efimov spectra of the
$^4$He tetramer states ($v=0,1$) associated with the 
trimer ground state 
using the same realistic $^4$He potentials as above
and  present that the calculated values of the 
$a_-^{(4,0)}$ and $a_-^{(4,1)}$ are independent of the  details
of the potentials at short inter-particle separations
and do not contradict the observed values 
when scaled with the van der Waals length. 
From the calculated tetramer wave function described 
with the full 18 sets of the Jacobi coordinates, 
we derive a four-body potential $U_4(R_4)$ as a function of the
four-body hyperradius $R_4$ and show that 
the repulsive core is located at $R_4 \approx 3 \, r_{\rm vdW}$ 
and the four-body probability density inside the core is 
heavily suppressed both in the ground and excited states.

The region suppressed by the three-(four-)body  barrier 
in the $^4$He trimer (tetramer) is significantly larger than the region
that is trivially excluded by the two-body repulsive core 
at $r \approx r_{\rm vdW}$, which makes the critical scattering lengths
insensitive to the short-range physics.

It is to be noted that the present work has the following limitation:
The realistic $^4$He-$^4$He potential with the van der Waals tail
is a single-channel potential into which
excitation of the interacting particles is renormalized,
and no excited channel is explicitly incorporated 
in the present framework. Therefore, we shall compare our results with
the property of the broad Feshbach resonances in the ultracold alkali 
atoms and consider that our single-channel model is not applicable 
to the narrow resonances generated by the multi-channel effects 
(cf. for instance, Refs.~\cite{Chin10,Wang14}). 

This paper is organized as follows: In Sec.~II, we briefly explain
the method and realistic $^4$He potentials.
In Sec.~III, we calculate the 
Efimov spectrum of the $^4$He trimer and derive the 
three-body parameters  $a_-^{(v)}$$(v=0,1)$ which are compared with the 
corresponding observed values.
The three-body hyperradial potentials $U_3^{(v)}(R_3)$ is derived. 
In Sec.~IV, we calculate the Efimov spectrum
of the $^4$He tetramer and derive  the critical scattering lengths
$a_-^{(4,v)}$$(v=0,1)$ which are compared with the observed values.
The four-body hyperradial potential $U_4^{(v)}(R_4)$ is derived. 
Summary is given in Sec.~V.

\section{METHOD AND INTERACTIONS}

Soon after the Efimov effect was predicted in the early 1970s,
the $^4$He trimer was expected 
to have bound states of Efimov type since 
the $^4$He-$^4$He interaction gives 
a large scattering length $a \sim 115$ \AA$\,$ (much greater 
than the potential range $\sim 10$ \AA)
and a very small $^4$He dimer binding energy $B_2 \sim 1$ mK
with the large mean radius $\langle r \rangle \sim 54$ \AA.
This fact strongly stimulated a large number of three-body calculations 
of $^4$He atoms, which predicted existence of the
ground state and a very weakly bound excited states 
(references in papers I and II as well as
the latest result for the $^4$He dimer, trimer and tetramer 
using realistic $^4$He potentials).

Experimentally, Ref.~\cite{Grisenti2000} obtained 
$\langle r \rangle=52 \pm 4$ \AA$\,$ and estimated 
$a=104^{+8}_{-18}$ \AA$\,$ and 
$B_2 =1.1^{+0.3}_{-0.2}$ mK using a crude model of
$a=2 \langle r \rangle$ and $B_2=\hbar^2/4 m \langle r \rangle^2$;
a more appropriate estimation 
was recently given in Ref.~\cite{PCKLJS-2} as 
$a=100.2^{+8.0}_{-7.9}\,$\AA$\,$ and $B_2= 1.30^{+0.25}_{-0.19}$ mK.
The $^4$He trimer ground state has been
observed  in Ref.~\cite{dimer-exp2005} 
to have the $^4$He-$^4$He bond length 
of $11^{+4}_{-5}$ \AA$\,$ in agreement with theoretical predictions,
whereas a reliable experimental evidence for the $^4$He trimer
excited state is still missing.

Although the Efimov trimers have been observed  
in experiments with ultracold alkali atoms,
the study of $^4$He clusters has been providing with 
the fundamental information to the universal Efimov physics.
In this paper, we intend to add the new understanding that
even $^4$He potentials are consistent with measurements 
in ultracold atoms for broad Feshbach resonances of
trimers and tetramers tied to the
trimer ground states and that this fact  
shows the large extent of universality in three- and four-body systems 
with interactions featuring van der Waals tails.

\subsection{Gaussian expansion method for few body systems}

In order to solve accurately the three- and four-body Schr\"{o}dinger 
equations for the $^4$He trimer and tetramer, we use 
the Gaussian expansion method for few-body 
systems~\cite {Kamimura88, Kameyama89, Hiyama03, Hiyama2012-FBS,
Hiyama2012-PTEP} explained precisely in papers I and II.
We employ the seven kinds of realistic 
$^4$He-$^4$He potentials;
names of the potentials are 
LM2M2~\cite {LM2M2}, TTY~\cite {TTY},  
HFD-B~\cite {HFD-B}, HFD-B3-FCI1~\cite {B3FCI1a, B3FCI1b, SAPT2},
SAPT96~\cite {SAPT96a, SAPT96b, SAPT2}, 
CCSAPT07~\cite {CCSAPT07}, and PCKLJS~\cite {PCKLJS,PCKLJS-2}. 
Among them the  PCKLJS  is currently 
the most sophisticated and accurate as explained in paper II,
while the LM2M2  has most popularly been used
in the literature.

The realistic $^4$He potentials has  
a strong repulsive core (Fig.~1), which makes it 
technically challenging to solve the low-energy 
few-body problems accurately. 
As was shown in papers I and II, however,
our method is suitable for describing both
the short-range correlations 
(without \mbox{{\it a priori}} \mbox{assumption} of any two-body 
correlation function) 
and the long-range asymptotic behavior of 
the $^4$He trimer and tetramer.

\begin{figure}[t]
\begin{center}
\epsfig{file=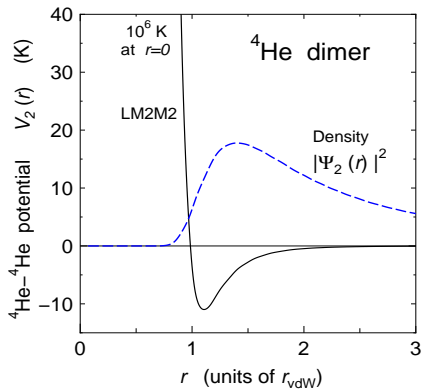,width=5.6cm,height=5.2cm}
\caption{(Color online) 
The LM2M2 potential $V_2(r)$, one of the realistic $^4$He potentials,
is shown as the black solid curve.  The dashed blue curve
shows the density of the dimer  $|\Psi_2 (r)|^2$
 (in arbitrary unit) at  $E_2=-1.309$ mK, so weakly bound compared 
with the potential pocket of $-11$ K. $\,r_{\rm vdW}=5.08 \,a_0, a_0$ 
being the Bohr radius.
}
\end{center}
\end{figure}
\begin{figure}[h]
\begin{center}
\epsfig{file=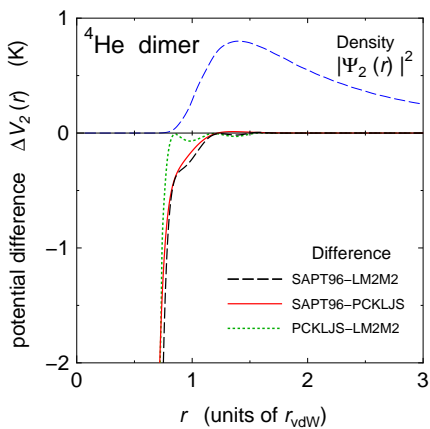,width=5.7cm,height=5.7cm}
\caption{(Color online) 
Difference in the realistic $^4$He potentials $V_2(r)$ 
between the three example potentials.
The dimer energy is  $E_2=-1.309$ mK(LM2M2),
$-1.615$ mK(PCKLJS) and $-1.744$ mK(SAPT96).
The repulsive core in the potentials 
is \mbox{$\approx\!1000$ K} at $r=0.7\, r_{\rm vdW}$.
The dashed blue curve
is the same density of the dimer in Fig.~1
 (in arbitrary unit).
}
\end{center}
\end{figure}

The wave function of the $A$-body $^4$He-atom system ($A=2 - 4$), 
say $\Psi_A$,  is obtained by solving 
the Schr\"{o}dinger equation in the fully nonadiabatic manner:
\begin{equation}
\big[\, T +  \sum_{1=i<j}^A \lambda \, V_2(r_{ij}) - E_A \, \big] \,\Psi_A =0 ,
\end{equation}
\noindent
where $T$ is the kinetic energy and 
$V_2(r_{ij})$ is the realistic  $^4$He potential
as a function of the pair separation $r_{ij}$ 
(see below for the factor $\lambda$).
$\Psi_A$ has the total angular momentum $J=0$.
The mass parameter is fixed to 
$\frac{\hbar^2}{m}=12.11928$ K\AA$^2$~\cite{newmass} where
$m$ is mass of  $^4$He atom.

The $^4$He tetramer wave function $\Psi_4$  
is calculated by expanding it in terms of the totally  
symmetrized \mbox{$L^2$-integrable} $K$-type and $H$-type four-body
Gaussian basis  functions of the full 18 sets of the Jacobi coordinates
as was explained in \mbox{paper I.}
The expansion coefficients and eigenenergy $E_4$ 
are determined by diagonalizing 
the Hamiltonian in the large function space. 
The nonlinear parameters, such as the Gaussian ranges
in  geometric progressions, taken in the calculation were 
explicitly tabulated in paper I.
Similar prescription is applied to the trimer (dimer) wave 
function $\Psi_3 (\Psi_2) $.

For plotting the three-(four-)body 
Efimov spectrum of the $^4$He trimer (tetramer), 
we calculate the  energies of the ground and excited states, 
$E_3^{(0)}$ and $E_3^{(1)}$ ($E_4^{(0)}$ and $E_4^{(1)}$), 
as functions of the
scattering \mbox{length $a$} of the two-body interaction.
Following the \mbox{literature~\cite{Esry96,Barletta01,
Gattobigio12,Naidon12}}, 
\mbox{we mimic} the experimental tunability of 
the interatomic interactions via broad 
Feshbach resonances by directly
altering the strength $\lambda$ of the realistic $^4$He potential
in Eq.~(2.1), which leads to the desired change  \mbox{of $a$}
(when $\lambda=1$, the potential equals 
to the original one). 

The realistic $^4$He potentials have the same
van der Waals potential $-C_6/r^6$ but the binding energy of the 
$^4$He dimer ranges from 1.309 mK (LM2M2) to 1.744 mK (SAPT96)
with some 30\% difference
(cf. \mbox{Table I} below)
due to the variation in the short-range physics. 
We illustrate in Fig.~2 
the potential differences between the three example potentials
(LM2M2, PCKLJ and SAPT96).
It is of interest to examine to what extent those
differences affect the critical
\mbox{scattering} lengths $a_-^{(v)}$ in the
$^4$He trimer and $a_-^{(4,v)}$ in the tetramer.  

\subsection{Three-(four-)body hyperradial potential}

In order to investigate the short-range repulsive barrier
in terms of the $A$-body hyperradius $R_A \:(A=3,4)$, 
we derive the hyperradial wave function from our wave 
function $\Psi_A^{(v)}$.
\mbox{Although} there are several conventions to define $R_A$,
the present paper employs the definition in 
Refs.~\cite{Fabre83,Gattobigio09}
for the $A$-body system with an equal mass:
\begin{equation}
R_A= \Big[\: \frac{2}{A} \sum_{j>i}^A ({\bf r}_i 
-{\bf r}_j)^2 \: \Big]^{1/2} ,  
\end{equation}
\noindent
where ${\bf r}_i$ is the position vector of 
particle $i$.

For the trimer, we have 
$R_3=\sqrt{\frac{2}{3}(r_{12}^2+r_{13}^2+r_{23}^2)}$
which was taken in the calculations in 
\mbox{Refs.~\cite{Efimov70,Gattobigio12,Endo12},}
while \mbox{Ref.~\cite{Wang12}} 
employed another definition of
\mbox{$R_3^{[18]}=
\sqrt{  \frac{1}{\sqrt{3}}(r_{12}^2+r_{13}^2+r_{23}^2)  } 
\approx 0.93R_3$.} 

We define the probability density of the $A$-body system, 
$\rho_A^{(v)}(R_A')\: (A=3,4 \; \mbox{and} \; v=0,1)$,
as a function of the hyperradius $R_A'$ by
\begin{equation}
\rho_A^{(v)}(R'_A) = 
\langle \,\Psi_A^{(v)}\,|\, \frac{\delta(R_A-R_A')}{R_A^{2D}}\,  | 
\,\Psi_A^{(v)}\,\rangle ,
\end{equation}
where  $D$ is given by $D=(3N-1)/2$ with 
\mbox{$N=A-1$~\cite{Fabre83}} and 
$\rho_A^{(v)}(R_A)$ is normalized
as $\int_0^\infty  \rho_A^{(v)}(R_A) R_A^{2D} dR=1$
since $\langle \,\Psi_A^{(v)}\,|\,\Psi_A^{(v)}\,\rangle=1$.
In Appendix, we explain in more detail how we calculate
Eq.~(2.3) using $\Psi_A^{(v)}$ which is written
in terms of the full sets of Jacobi coordinates.

We relate the hyperradial density $\rho_A^{(v)}(R_A)$ 
to the \mbox{hyperradial} wave function $f_A^{(v)}(R_A) \:(v=0,1)$
by 
\begin{equation}
\rho_A^{(v)}(R_A) = \left| \frac{f_A^{(v)}(R_A)}{R_A^D} \right|^2 ,
\end{equation}
\noindent
and consider that
$f_A^{(v)}(R_A)$ satisfies the single-channel
hyperradial Schr\"{o}dinger equation for the $^4$He trimer
and for the tetramer associated with the ground-state trimer:
\begin{equation}
\Bigg[ -\frac{\hbar^2}{m} \frac{d^2}{dR_A^2}
  + U_A(R_A) -E_A^{(v)} \, \Bigg] f_A^{(v)}(R_A)=0 ,
\end{equation}
(cf. Eq.~(4.4) in Ref.~\cite{Fabre83}, Eq.~(6) in Ref.~\cite{Endo12}
and Eq.~(1) in Ref.~\cite{Wang12}). 
In Eq.~(2.5), the nonadiabatic effect 
and the coupling to other channels appearing 
in the hyperspherical framework
are all renormalized into the $A$-body hyperradial 
potential $U_A(R_A)$ since $f_A^{(v)}(R_A) \:(v=0,1)$
is calculated from the fully nonadiabatic solution  $\Psi_A^{(v)}$ 
and $E_A^{(v)}$ of the original Schr\"{o}dinger equation (2.1).

We  derive $U_A^{(v)}(R_A)\:(v=0,1)$ by
\begin{equation}
     U_A^{(v)}(R_A) =  \frac{\hbar^2}{m} 
\frac{d^2 f_A^{(v)}(R_A)}{dR_A^2} \Big/  {f_A^{(v)}(R_A)} - E_A^{(v)} .
\end{equation}
The result should satisfy 
$ U_A^{(0)}(R_A) =  U_A^{(1)}(R_A)$ as the potential
$ U_A(R_A)$  in the
Schr\"{o}dinger Eq.~(2.5); this will successfully be examined in
Secs.~III$\,$B and IV$\,$B.  

At large $R_3$, one would expect that  
$U_3(R_3)$ asymptotes to
\begin{equation}
U_3(R_3) \to - \frac{\hbar^2}{m} \frac{s_0^2 + 
\frac{1}{4}}{R_3^2} \quad (R_3 \gg r_{\rm vdW})
\end{equation}
with $s_0 \approx 1.00624$  as the usual 
Efimov behavior of the potential~\cite{Efimov70,Wang12,Endo12}.
It will be shown in Sec.~III$\,$B that 
$U_3^{(v)}(R_3)\; (v=0,1)$ satisfy Eq.~(2.7).

We note that if the two-body repulsive core at 
\mbox{$r_{12} \approx r_{\rm vdW}$}  is the direct  origin of 
the repulsive  barrier in $U_A(R_A)$,
the particles could come close together reaching 
\mbox{$R_A \approx \sqrt{A-1}\, r_{\rm vdW}$}  
when all the \mbox{$r_{ij} \approx r_{\rm vdW}$} in Eq.~(2.2);
but this  will be denied in Secs.~III$\,$B and IV$\,$B
since the possible minimum $R_A$ is derived as 
\mbox{$R_A \approx (A-1)\, r_{\rm vdW}\: (A=3,4)$}
in the actual calculation.

\section{Result for $^4$H\lowercase{e} trimer}

\subsection{Three-body parameter}

The calculated Efimov spectrum for the $^4$He trimer is
plotted in Fig.~3 with the solid curves; 
the trimer energies $E_3^{(0)}$ and $E_3^{(1)}$ are illustrated
as  functions of the scattering length $a$.
Following the literature, we have drawn 
$(|E|/E_{\rm vdW})^{1/4}$  versus
$(|a|/r_{\rm vdW})^{-1/2}$  so that both curves are 
graphically represented on the same scale.
The scattering length $a$ and the energy $E$ are scaled
with $r_{\rm vdW}\, (=5.08\, a_0$~\cite{Chin10}) 
and the van der Waals energy
$E_{\rm vdW}=\hbar^2/m r_{\rm vdW}^2\, (=1.677$ K), respectively.
The dashed curve shows the dimer energy.
The vertical dotted line indicates 
the physical value $\lambda=1$.

The result in Fig.~3 depends little on the seven kinds of the
realistic potentials so that
the curves  are the same for different potentials 
within the thickness of the lines.
We note that, in the case of the LM2M2 potential, 
almost the same figure as Fig.~3
was already reported by one of the authors (E.H.) and 
the collaborators~\cite{Naidon12} 
and by Gottobigio {\it et al.}~\cite{Gattobigio12}.

In Fig.~3, $a_-^{(0)}$ and $a_-^{(1)}$ indicate the critical scattering 
lengths at which the trimer energies $E_3^{(0)}$ and $E_3^{(1)}$ 
intersect respectively the three-atom threshold on the $a<0$ side.
The calculated values of them are summarized 
in \mbox{Table~I} together with  the recent experimental values 
for the alkali-atom trimers~\cite{Kraemer06,Pollack09,Gross09,Wenz09,
Ferlaino09,Williums09,Gross10,Berninger11,
Wild12,Dyke13,Zenesini13,Huang14}
except for those on narrow Feshbach resonances.  
First of all, we emphasize that the seven realistic potentials
give the same value as 
\mbox{$a_-^{(0)}/a_0=-48.2(1)$} and \mbox{$a_-^{(1)}/a_0=-832(1)$} 
showing a negligible contribution from the difference in their
short-range details shown in Fig.~2.

\begin{figure}[b]
\begin{center}
\epsfig{file=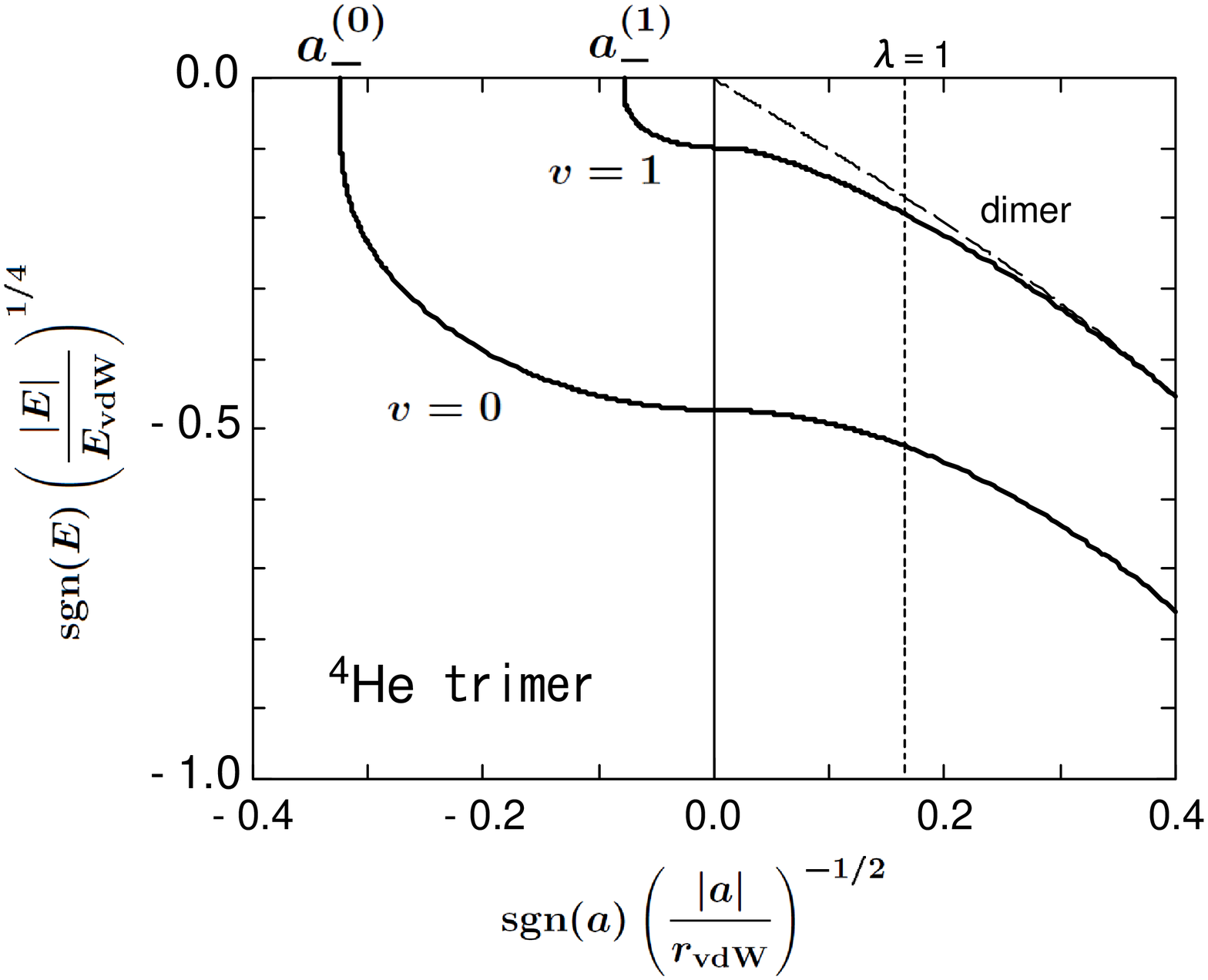,width=8.0cm,height=6.5cm}
\caption{ 
Efimov spectrum for the $^4$He trimer calculated 
with the seven realistic $^4$He potentials,
LM2M2, TTY, HFD-B, HFD-B3-FCI1, SAPT96, 
CCSAPT07, and PCKLJS: trimer energy $E_3^{(v)}$ 
as a function of the scattering length $a$
for the ground $(v=0)$ and excited $(v=1)$ states 
are shown as the solid curves, while
the dimer energy is shown as the dotted curve.
The curves for different potentials overlap with
each other within the line thickness.
To clarify the figure, $a$ and $E$ are scaled with the 
$r_{\rm vdW}$ and $E_{\rm vdW}$, and
raised to the power $-1/2$ and 1/4, respectively.
$a_-^{(0)}$ and $a_-^{(1)}$ denote 
respectively the scattering lengths where 
the trimer energies $E_3^{(0)}$ and $E_3^{(1)}$ intersect the 
three-atom threshold.
The dotted line indicates the scattering length 
corresponding to the unscaled potential at $\lambda=1$.
}
\end{center}
\end{figure}
\begin{table*}[t]
\caption{Scattering lengths for the first and second Efimov resonances, 
$a_-^{(0)}$ and $a_-^{(1)}$, where the Efimov spectra $(v=0,1)$ cross 
the three-body threshold on the negative $a$ side (cf. Fig.~3).
They are calculated using the seven different realistic $^4$He potentials
(see text for $\lambda$ and $\tilde{r}_{\rm vdW}$). 
The values scaled with $\tilde{r}_{\rm vdW}$ are 
compared with experimental values scaled with $r_{\rm vdW}$.
Here, $r_{\rm vdW}/a_0= 5.08,$ $\,$ 31.26, $\,$ 32.49, $\,$ 82.10 and 101
for $^4$He, $^6$Li, $^7$Li, $^{85}$Rb and $^{113}$Cs, 
respectively~\cite{Chin10}.
The potential names are 
arranged in the increasing order of the binding 
energy of the $^4$He dimer, $B_2$,  at $\lambda=1$.
}
\begin{center}
\begin{tabular}{lcccccccccc} 
\hline \hline
\noalign{\vskip 0.1 true cm} 
$\; ^4$He $\,$ trimer    & \qquad  & 
              \multicolumn{3}{c} { Ground state $(v=0)$}  &    &
              \multicolumn{3}{c}  { Excited state $(v=1)$ }    \\
\noalign{\vskip -0.2 true cm} 
   & \qquad    &  \multispan3 {\hrulefill} 
            &   & \multispan3 {\hrulefill} \\
\noalign{\vskip 0.01 true cm} 
   Realistic potentials & $\;B_2$ (mK)$\;\;\;\;\;\;$  & $a_-^{(0)}/a_0$     &  
               $a_-^{(0)}/\tilde{r}_{\rm vdW}$  
              &   $\lambda$ &  & $a_-^{(1)}/a_0$ \qquad    &      
               $a_-^{(1)}/\tilde{r}_{\rm vdW}$   
              & $\lambda$   \\ 
\noalign{\vskip 0.1true cm} 
\hline 
\noalign{\vskip 0.1 true cm} 
LM2M2  & 1.309 $\;\;$& 
     $\;$ $-48.26$\footnotemark[1]  &  $-9.78$    
     & $\;$ 0.8901$\;\;$ & \quad \quad  & 
     $\;$ $-832$\footnotemark[1]   &   $-165$ 
     & $\;$ 0.9685   \\ 
TTY  & 1.316 $\;\;$& 
      $-48.29$ &  $-9.78$  &  0.8902 &   & 
      $-832$  &   $-165$ &     $\;$ 0.9685   \\ 
HFD-B3-FCI1  & 1.448 $\;\;$& 
      $-48.22$ &  $-9.78$  &    0.8891 &   & 
      $-832 $  &   $-165$ &     $\;$ 0.9673  \\ 
CCSAPT07  & 1.564 $\;\;$& 
      $-48.21$ &  $-9.78$  &    0.8881 &   & 
      $-832 $  &   $-165$ &    $\;$ 0.9662   \\ 
PCKLJS   & 1.615 $\;\;$& 
      $-48.21$ &  $-9.78$  &    0.8877 &   & 
      $-832 $  &   $-165$ &    $\;$ 0.9658   \\ 
HFD-B   & 1.692 $\;\;$& 
      $-48.22$ &  $-9.78$  &    0.8869 &   & 
      $-832 $  &   $-165$ &    $\;$ 0.9651    \\ 
SAPT96  & 1.744 $\;\;$& 
      $-48.18$ &  $-9.77$  &    0.8867 &   & 
      $-831 $  &   $-165$ &    $\;$  0.9647   \\ 
\noalign{\vskip 0.1 true cm} 
\hline
\noalign{\vskip 0.1 true cm} 
\noalign{\vskip 0.01 true cm} 
   Experiments &   & $a_-^{(0)}/a_0$     &  
               $a_-^{(0)}/r_{\rm vdW}$  &   & 
             & $a_-^{(1)}/a_0$ \qquad    &      
               $a_-^{(1)}/r_{\rm vdW}$   &    \\ 
\noalign{\vskip 0.1 true cm} 
\hline
\noalign{\vskip 0.1 true cm} 
Exp($^{133}$Cs, $^{85}$Rb,$^{6,7}$Li )    &  & 
       &  $\approx -9.5\pm 15 \%$\footnotemark[2]   &  
          &   & 
       &    & \quad$\,$      \\ 

Exp ($^{133}$Cs)~\cite{Huang14}  & \quad & 
        $-963(11)$  &   $-9.53(11)$   &   & 
      &$-20190 (1200)$   
    &  $\;\;\;\;-200(12)$ & \quad$\,$       \\ 

%

Exp ($^6$Li)~\cite{Wenz09,Williums09}  &  & 
      $-292$   & $-9.34$  &   &   & 
      $-5752$   &   $-177$
        & \quad$\,$    \\ 
\noalign{\vskip 0.1 true cm} 
\hline
\hline
\noalign{\vskip -0.3 true cm} 
\end{tabular}
\label{table:trimer}
\end{center}
\footnotetext[1]{Ref.~\cite{Gattobigio12} gave 
$a_-^{(0)}/a_0 \sim -48.1$
and $a_-^{(1)}/a_0 \sim -975$ for LM2M2.}
\footnotetext[2]{A value summarized 
in ~\cite{Berninger11,Machtey12,Wang12,Roy13}
for experimental data~\cite{Kraemer06,Pollack09,Gross09,Wenz09,
Ferlaino09,Williums09,Gross10,Berninger11,Wild12,Dyke13,Zenesini13,
Huang14}.}
\end{table*}

Here, we introduce a modified van der Waals length, 
say $\tilde{r}_{\rm vdW}$,
for the scaled $^4$He potential $\lambda V_2(r_{ij})$
that gives  $a_-^{(0)}$ and $a_-^{(1)}$ with the $\lambda$ values
listed in Table I:
\begin{equation}
\tilde{r}_{\rm vdW}=\frac{1}{2}
(m \lambda C_6 /\hbar^2)^{\frac{1}{4}}
=\lambda^{\frac{1}{4}} r_{\rm vdW}.
\end{equation}
The calculated and observed values of $a_-^{(0)}$ and $a_-^{(1)}$   
are scaled by $\tilde{r}_{\rm vdW}$
for $^4$He and by $r_{\rm vdW}$ for alkali atoms 
and are compared in Table I (the same applies to tetramers). 
For the $^4$He trimer,   
$\lambda^{\frac{1}{4}}\!\approx\! 0.97$ for $a_-^{(0)}$ and
$\approx\!0.99$ for $a_-^{(1)}$. 
 
The calculated value of 
$a_-^{(0)}/\tilde{r}_{\rm vdW}=9.78(1)$ 
agrees with the universal value  
$a_-^{(0)}/r_{\rm vdW} \approx -9.5\pm 15 \%$
obtained in the recent experiments and  
$a_-^{(0)}/r_{\rm vdW} = -9.73(3)\pm 15\%$ 
calculated in Ref.~\cite{Wang12},
while $a_-^{(1)}/\tilde{r}_{\rm vdW}=165(1)$ 
for the excited state of the $^4$He
trimer~\footnote{ 
It is hard to calculate precisely $a_-^{(1)}$ of the
excited trimer state that  has an extremely large size
at  $E_3^{(1)}\approx 0$.
We took the maximum range of the Gaussians to be 4,000 $r_{\rm vdW}$ 
with an increased number of basis when calculating 
$a_-^{(v)} (a_-^{(4,v)})$.
We note that, if  a smaller-size (worse) three-body
basis set is {\it artificially} employed, 
the solid curve ($v=1$) in Fig.~3 
shifts slightly to the right (namely, the eigenstate 
becomes shallower), which gives 
a larger value of $|a_-^{(1)}|/\tilde{r}_{\rm vdW}$,    
even {\it closer} to the corresponding observed data for the
alkali atoms.
}  
does not contradict the
observed values $a_-^{(1)}/r_{\rm vdW}=-177$ and $-200(12)$ 
for the second Efimov resonances in $^{6}$Li and $^{133}$Cs, 
respectively.

The low-energy universal Efimov theory does not predict the absolute
position of $a_-^{(0)}$, but predicts the ratio $a_-^{(1)}/a_-^{(0)}$ to
have a universal value $a_-^{(1)}/a_-^{(0)}
=22.7$~\cite{Efimov70,Braaten06,Braaten07}. According to the
literature~\cite{DIncao09,Platter09,Thogersen08,Schmidt12,Wang14},
effects of the finite interaction range and the van der
Waals tail give corrections toward somewhat smaller values than 22.7;
especially, a value of 17.1 was predicted in Ref.~\cite{Schmidt12}
in the limit of strongly entrance-channel-dominated 
Feshbach resonances. The experiments
that measured both the $a_-^{(0)}$ and $a_-^{(1)}$ in the same atom gave
the ratio  $(a_-^{(1)}/a_-^{(0)})_{\rm exp}=21.0(1.3)$ and $19.7$ for
$^{133}$Cs~\cite{Huang14} and $^6$Li~\cite{Williums09,Wenz09}
respectively. In the present 
calculation of the $^4$He trimer using the realistic $^4$He potentials, 
we have also found that the ratio gets smaller than the universal value,
giving $(a_-^{(1)}/a_-^{(0)})_{^4{\rm He}}=17.2$~\footnote{  
The same as the comments on  $|a_-^{(1)}|/r_{\rm vdW}$ in [61]
for the ratio  $a_-^{(1)}/a_-^{(0)}$.
}.    

\subsection{Three-body repulsive barrier}

Following the formulation of the hyperradial three-body potential in
Sec.~II$\,$B, we first calculate the wave function $\Psi_3^{(v)}$ 
in Eq.~(2.1) using  
the LM2M2 potential, as an example (use of the other potentials give 
almost the same result), 
at the unitary limit $(|a|\to\infty)$ given by 
$\lambda=0.9743$. We obtain  $E_3^{(0)}=-0.0501\, E_{\rm vdW}$
and $E_3^{(1)}=-9.36 \times 10^{-5}\, E_{\rm vdW}$; 
the r.m.s. radius $\langle R_3^2 \rangle^{1/2}$ is
6.47 $r_{\rm vdW}\: (v=0)$ and 94.2 $r_{\rm vdW}\: (v=1)$. 
The ratio $E_3^{(0)}/E_3^{(1)}=534$ is close to the ratio 531 in 
Ref.~\cite{Schmidt12} in the case of a broad Feshbach resonance with the 
strength parameter $s_{\rm res}=100$. This again implies the applicability 
of the present calculation using the single-channel atom-atom potential 
to systems featuring broad Feshbach resonances.

Calculated  three-body densities $|f_3^{(v)}(R_3)|^2$ are illustrated
in Fig.~4 for $R_3< 7 \,r_{\rm vdW}$ and 
in Fig.~5 (log scale) for \mbox{$R_3< 25 \, r_{\rm vdW}$} 
as the solid curve $(v=0)$ and the dotted red curve ($v=1$). 
In Fig.~6, the three-body potentials $U_3^{(v)}(R_3)$ defined by Eq.~(2.6)
are shown as the solid curve $(v=0)$ and the
dotted red curve ($v=1$). 
Also shown,  for the sake of reference,
are the Efimov attraction, Eq.~(2.7), in
the dot-dash curve and the density $|f_3^{(0)}(R_3)|^2$ in the 
dashed blue curve (arbitrary unit).

\begin{figure}[t]
\begin{center}
\epsfig{file=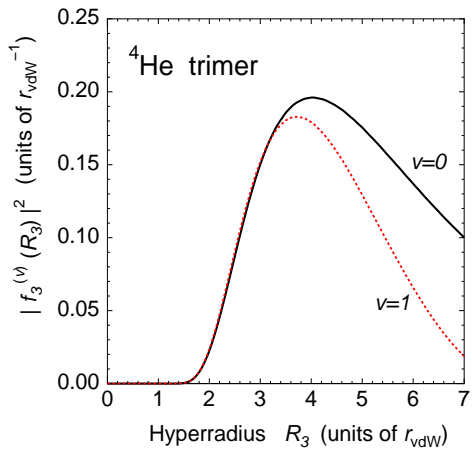,width=6.5cm,height=6.0cm}
\caption{(Color online) 
Probability density $|f_3^{(v)}(R_3)|^2 \:(v=0,1)$ of the $^4$He trimer 
ground (solid curve) and excited  (dotted red curve) states
versus the three-body hyperradius $R_3$ for \mbox{$R_3< 7\,r_{\rm vdW}$.}
The dotted curve has been multiplied by a factor 463 to emphasize
that  the two states exhibit the same shape of 
the strong short-range correlations 
for $R_3 \lesssim 3\, r_{\rm vdW}$.
} 
\end{center}
\end{figure}
%
\begin{figure}[t]
\begin{center}
\epsfig{file=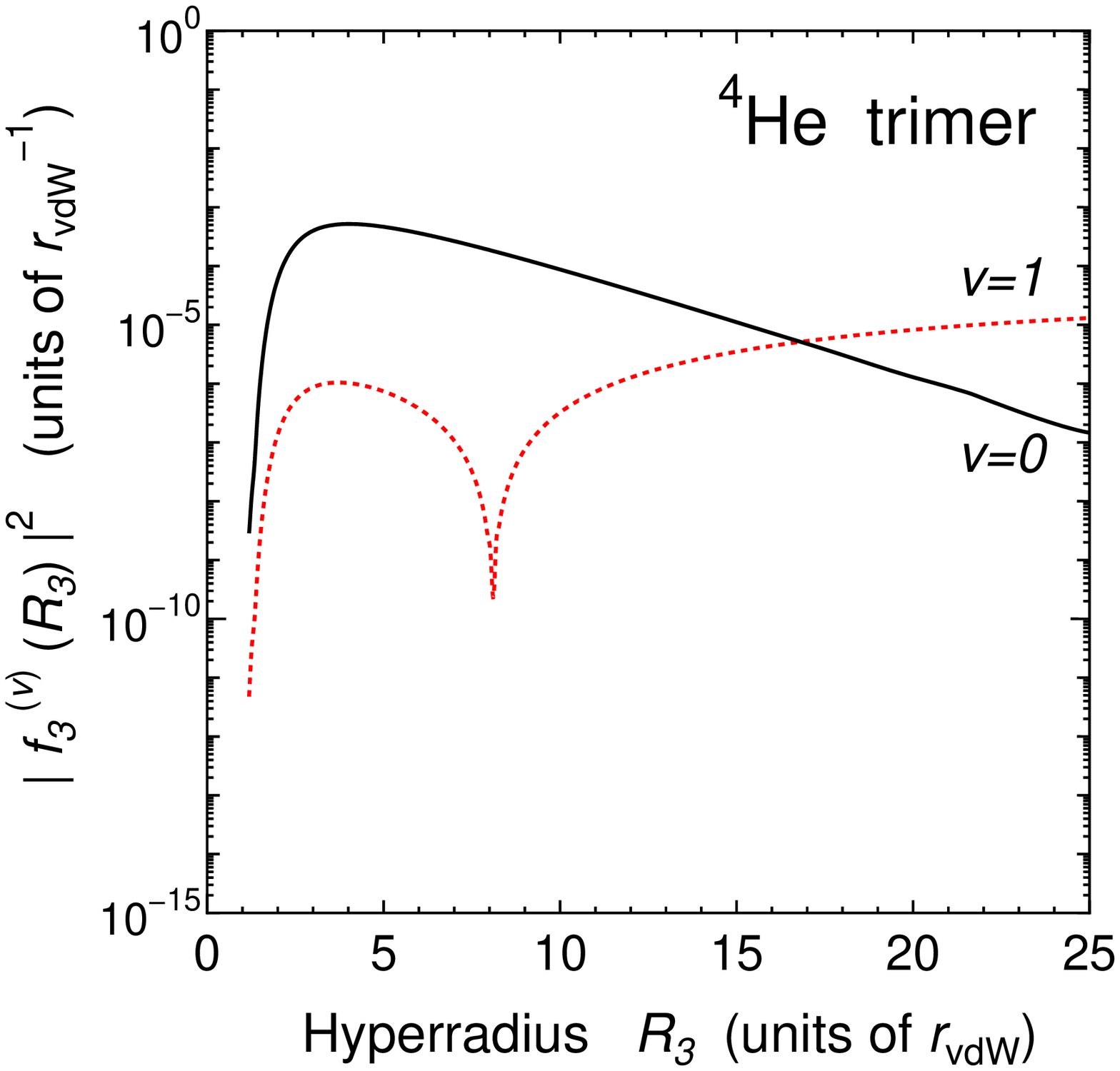,width=6.2cm,height=6.2cm}
\caption{(Color online) 
The same as Fig.~4 for $R_3 < 25\,r_{\rm vdW}$ in the log sale,
but no factor is multiplied to the dotted curve.
}
\end{center}
\end{figure}

\begin{figure}[t]
\begin{center}
\epsfig{file=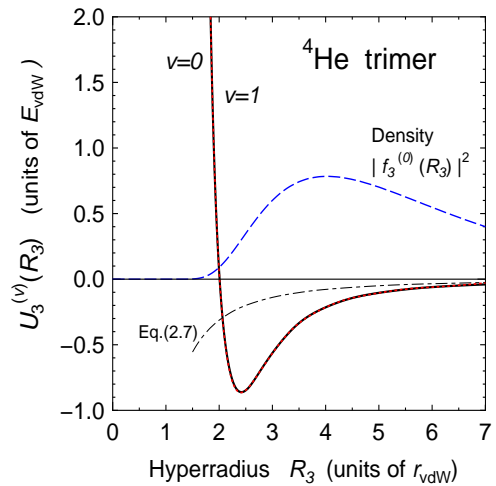,width=6.5cm,height=6.5cm}
\caption{(Color online) 
The three-body potential $U_3^{(v)}(R_3)\;(v=0,1)$ defined by
Eq.~(2.6) for the trimer ground (solid curve) and excited (dotted red curve)
states. The two curves overlap with each other within the thickness
of the lines. The dash-dotted curve shows 
the Efimov attraction, Eq.~(2.7), 
and the dashed blue curve is for the density $|f_3^{(0)}(R_3)|^2$  
in arbitrary unit.
}
\end{center}
\end{figure}

A striking aspect of these figures is that the hyperradial potentials
satisfy $U_3^{(0)}(R_3)=U_3^{(1)}(R_3)$  and 
converge to the Efimov attraction at large distances;
this demonstrates the validity of the idea
to construct the hyperradial Schr\"{o}dinger equation (2.5)
starting from the solution $\Psi_3^{(v)}$ and $E_3^{(v)}$ of the
original Schr\"{o}dinger equation (2.1). 

From the behavior of the potentials $U_3^{(v)}(R_3)$  
and the densities $|f_3^{(v)}(R_3)|^2 \; (v=0,1)$, we recognize that
\mbox{there is} a three-body repulsive barrier 
at $R_3 \approx 2\,r_{\rm vdW}$.
Inside the barrier, $R_3 \lap 2\,r_{\rm vdW}$, 
the probability of finding the atoms is heavily suppressed.
This result supports the finding, in Refs.\cite{Wang12,Endo12}, 
of the appearance of the effective three-body repulsion with 
the adiabatic hyperspherical calculations.
Our potential in Fig.~6 is close to the
the hyperradial three-body potential for the first 
Efimov trimer state
obtained in Fig.~3b of Ref.~\cite{Wang12} 
and Figs.~1 and 7 of Ref.~\cite{Endo12}.

The  hyperradius 
$R_3 \approx 2\,r_{\rm vdW}$ of the three-body repulsive barrier
corresponds
to the inter-particle distance $r_{ij} \approx \sqrt{2} \,r_{\rm vdW}$ 
when all the $r_{ij}$ are equal to each other in Eq.~(2.2).  
The distance
is $\sqrt{2}$ times larger than the radius of the two-body potential core  
$r_{12} \approx r_{\rm vdW}$ and is located almost outside the region
where the difference in the seven realistic potentials is seen 
in Fig.~2; this is due to the nonadiabatic three-body dynamics that, 
as was pointed out in Refs.~\cite{Wang12,Endo12}, the suppression
of the two-body probability for $r_{ij} \lap r_{\rm vdW}$
leads to the three-body repulsion for $R_3 \lap 2\, r_{\rm vdW}$.

\section{Result for $^4$H\lowercase{e} tetramer}

\subsection{Four-body Efimov spectrum}

Using the same method of papers I and  II, we solved 
the four-body Schr\"{o}dinger equation (2.1)  for
the ground $(v=0)$ and excited $(v=1)$ states of the
$^4$He tetramer changing the factor $\lambda$ for the potentials.
The four-body Efimov spectrum of $E_4^{(0)}$ and $E_4^{(1)}$ 
is plotted in Fig.~7 in the solid curves 
together with the spectrum of $^4$He trimer in the 
thin solid blue curves.

We note that the four-body Efimov plot for the $^4$He tetramer
calculated with the {\it realistic} $^4$He potential 
is reported for the first time in the present paper.
Using effective two-body plus three-body Gaussian potentials,
Gattobigio {\it et. al}~\cite{Gattobigio12}
obtained a similar plot as Fig.~7 for the $^4$He tetramer. 
We find two tetramer bound states, 
one deep and one shallow, tied to the trimer ground state;
this agrees with the key prediction 
in Refs.~\cite{Hammer07,Stecher09}
that there are two universal four-body states associated with 
each Efimov trimer.

\begin{figure}[t]
\begin{center}
\epsfig{file=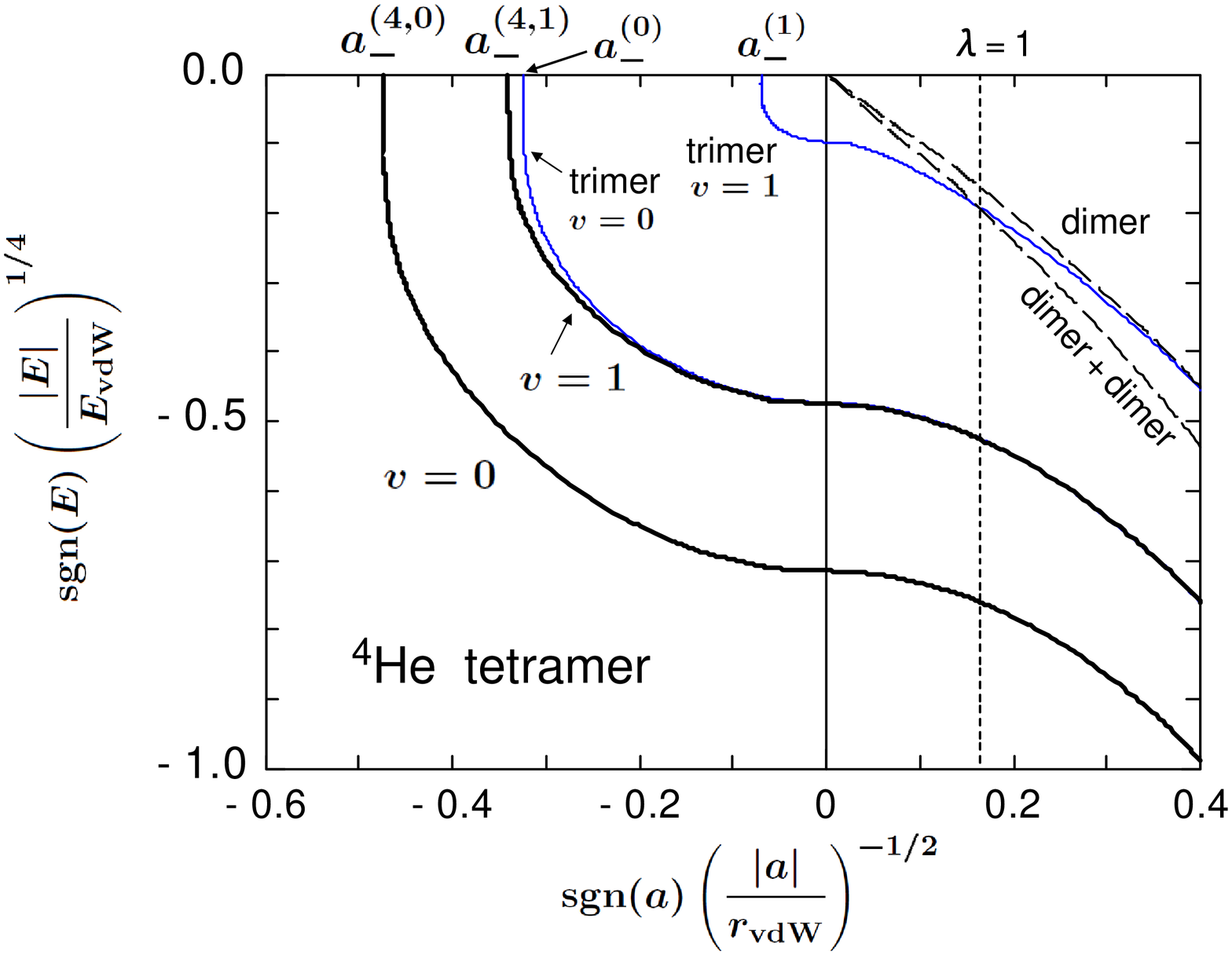,width=8.5cm,height=6.5cm}
\caption{(Color online) 
Efimov spectrum for the $^4$He tetramer calculated 
with the seven realistic $^4$He potentials,
LM2M2, TTY, HFD-B, HFD-B3-FCI1, SAPT96, 
CCSAPT07, and PCKLJS: The scaled tetramer energy $E_4^{(v)}/E_{\rm vdW}$ 
as a function of the scaled-inverse scattering length 
$(a/r_{\rm vdW})^{-1}$
for the ground $(v=0)$ and excited $(v=1)$ states.
The curves for different potentials overlap with
each other within the line thickness.
The thin solid blue curves denote  the trimer spectrum 
which is the same as that in Fig.~3.
$a_-^{(4,0)}$ and $a_-^{(4,1)}$  denote 
the critical scattering lengths where 
the tetramer energies $E_4^{(0)}$ and $E_4^{(1)}$ cross the 
four-atom threshold, respectively.
}
\end{center}
\end{figure}

\begin{table*}[t]
\caption{The critical scattering lengths  
$a_-^{(4,0)}$ and $a_-^{(4,1)}$, where the tetramer energies
$E_4^{(0)}$ and $E_4^{(1)}$ cross
the four-body threshold on the negative $a$ side (see Fig.~7).
They are calculated with the seven different realistic $^4$He potentials
(see text for $\lambda$ and $\tilde{r}_{\rm vdW}$). 
The values scaled with $\tilde{r}_{\rm vdW}$ are 
compared with experimental values scaled with $r_{\rm vdW}$.
$B_2$ is the binding energy of the dimer at $\lambda=1$.
}
\begin{center}
\begin{tabular}{lcccccccccc} 
\hline \hline
\noalign{\vskip 0.1 true cm} 
$\; ^4$He $\,$ tetramer    & \qquad  & 
              \multicolumn{3}{c} { Ground state $(v=0)$}  &    &
              \multicolumn{3}{c}  { Excited state $(v=1)$ }    \\
\noalign{\vskip -0.2 true cm} 
   & \qquad    &  \multispan3 {\hrulefill} 
            &   & \multispan3 {\hrulefill} \\
\noalign{\vskip 0.01 true cm} 
   Realistic potentials & $\;B_2$ (mK)$\;\;\;\;\;\;$  & $a_-^{(4,0)}/a_0$  & 
               $a_-^{(4,0)}/\tilde{r}_{\rm vdW}$ 
              &   $\lambda$ & 
             & $a_-^{(4,1)}/a_0$ \qquad    &      
           $\;\;$    $a_-^{(4,1)}/\tilde{r}_{\rm vdW}$  
             & $\lambda$   \\ 
\noalign{\vskip 0.1true cm} 
\hline 
\noalign{\vskip 0.1 true cm} 
LM2M2  & 1.309 $\;\;$& 
     $-22.69$  &  $-4.69$    
     & $\;$ 0.8197$\;\;$ & \quad \quad  & 
     $-43.98$   &   $-8.93$   & $\;$ 0.8832   \\ 
TTY  & 1.316 $\;\;$& 
      $-22.70$ &  $-4.69$  &    0.8199 &   & 
      $-43.96$  &   $-8.93$ &     $\;$ 0.8832   \\ 
HFD-B3-FCI1  & 1.448 $\;\;$& 
      $-22.67$ &  $-4.69$  &    0.8187 &   & 
      $-43.95 $  &   $-8.93$ &     $\;$ 0.8821  \\ 
CCSAPT07  & 1.564 $\;\;$& 
      $-22.66$ &  $-4.69$  &    0.8178 &   & 
      $-43.95 $  &   $-8.93$ &    $\;$ 0.8812   \\ 
PCKLJS   & 1.615 $\;\;$& 
      $-22.66$ &  $-4.69$  &    0.8174 &   & 
      $-43.93 $  &   $-8.93$ &    $\;$ 0.8807   \\ 
HFD-B   & 1.692 $\;\;$& 
      $-22.67$ &  $-4.69$  &    0.8166 &   & 
      $-43.95 $  &   $-8.93$ &    $\;$ 0.8800    \\ 
SAPT96  & 1.744 $\;\;$& 
      $-22.65$ &  $-4.69$  &    0.8165 &   & 
      $-43.91 $  &   $-8.92$ &    $\;$  0.8798   \\ 
\noalign{\vskip 0.1 true cm} 
\hline
\noalign{\vskip 0.1 true cm} 
Experiments &  & $a_-^{(4,0)}/a_0\;\;$  & 
               $\;\;a_-^{(4,0)}/r_{\rm vdW}$  &    & 
             & $a_-^{(4,1)}/a_0$ \qquad    &      
            $\;\;$       $a_-^{(4,1)}/r_{\rm vdW}$   &    \\ 
\noalign{\vskip 0.1true cm} 
\hline 
\noalign{\vskip 0.1 true cm} 
Exp ($^{133}$Cs)~\cite{Ferlaino-FB11}  & \quad & 
      $-444(8)$  & $\;\;\;\;-4.40(8)$  & &
     &$-862(9)$  &   $\;\;\;\;-8.53(9)$  &     \\ 

Exp ($^{133}$Cs)~\cite{Ferlaino09}  & \quad &  
      $-410\;\;\;\;$ &  $\;\;\;\;-4.06\;\;\;\;\;$  &  & 
     &$-730\;\;\;\;$  &   $\;\;-7.23\;\;$  &  \\ 

Exp ($^{133}$Cs)~\cite{Zenesini13}  & \quad & 
      $-440(10)$ &  $\;\;\;\;-4.36(10)$  & &
       & &   &   \\ 

Exp ($^7$Li)~\cite{Dyke13}  & \quad & 
      $-94(4)$ &  $\;\;\;\;-2.9(1)$  & &
     & $-236(10)$  &  $\;\;\;\;\;-7.26(31)$ &  \\ 

\noalign{\vskip 0.1 true cm} 
\hline
\hline
\noalign{\vskip -0.3 true cm} 
\end{tabular}
\label{table:trimer}
\end{center}
\end{table*}

In Fig.~7,  $a_-^{(4,0)}$ and $a_-^{(4,1)}$  denote the critical
scattering lengths where 
the tetramer energies $E_4^{(0)}$ and $E_4^{(1)}$ cross
the four-atom threshold.
Their values are summarized in \mbox{Table II} for the realistic 
$^4$He potentials together with the observed values for
\mbox{$^{133}$Cs~\cite{Ferlaino09,Ferlaino-FB11,Zenesini13}} and
\mbox{$^7$Li~\cite{Dyke13}}. 
Similarly to Table~I for the trimer,
the seven realistic potentials give the same value as
\mbox{$a_-^{(4,0)}/a_0=-22.7(1)$} 
and \mbox{$a_-^{(4,1)}/a_0=-44.0(1)$} showing
that they are insensitive to the details of the potentials
at short distances~\footnote{   
Using effective two-body plus three-body Gaussian potentials
without the van der Waals tail,
Ref.~\cite{Gattobigio12} gave
\mbox{$a_-^{(4,0)}/a_0=-19.6$} and $a_-^{(4,1)}/a_0=-39.8$,
respectively.
Their absolute values are 
by some $10-15$\% smaller than those in the present work
probably due to the lack of the long-range tail; the same 
tendency is seen in the trimer.
}.   
The calculated values of 
\mbox{$a_-^{(4,0)}/\tilde{r}_{\rm vdW}=-4.69(1)$} 
and \mbox{$a_-^{(4,1)}/\tilde{r}_{\rm vdW}=-8.93(1)$}
do not contradict the corresponding 
observed values for the 
$^{133}$Cs and $^7$Li tetramers as seen in \mbox{Table II.}

Reference~\cite{Stecher09} predicted that
the universal properties of the four-body system
are directly related to the three-body subsystem as 
$a_-^{(4,0)}=0.43 \, a_-^{(0)}$ and 
\mbox{$a_-^{(4,1)}=0.90 \, a_-^{(0)}$.} The present calculation of the 
$^4$He atoms using the realistic potentials 
gives consistent values of $a_-^{(4,0)} = 0.47 \, a_-^{(0)}$ 
and $a_-^{(4,1)}=0.91 \, a_-^{(0)}$.
The observed data for $^{133}$Cs and $^7$Li are in accordance with 
the prediction~\cite{Stecher09}.

\subsection{Four-body repulsive barrier}

Using the nonadiabatic solution
$\Psi_4^{(v)} \; (v=0,1)$ to the four-body Schr\"{o}dinger
equation (2.1), we calculate the probability density 
$|f_4^{(v)}(R_4)|^2$ and the  potential $U_4^{(v)}(R_4)$,
defined in Sec.~II,  as functions of
the four-body \mbox{hyperradius} 
\mbox{$R_4 =\sqrt{\frac{1}{2}(r_{12}^2 +r_{13}^2 +r_{14}^2 +r_{23}^2 
+r_{24}^2 + r_{34}^2 )}$.}
As for the $^4$He realistic potential, we employ, 
the LM2M2 potential, as an example,
at the unitary limit $(|a|\to\infty)$ with 
$\lambda=0.9743$; we have 
$E_4^{(0)}=-0.262 \: E_{\rm vdW}$ 
and $E_4^{(1)}=-0.0510 \: E_{\rm vdW}$.
The \mbox{r.m.s. hyperradius} $\langle R_4^2 \rangle^{1/2}$ is
5.95 $r_{\rm vdW}\: (v=0)$ and \mbox{38.0 $r_{\rm vdW}\:$} $(v=1)$.

\begin{figure}[t]
\begin{center}
\epsfig{file=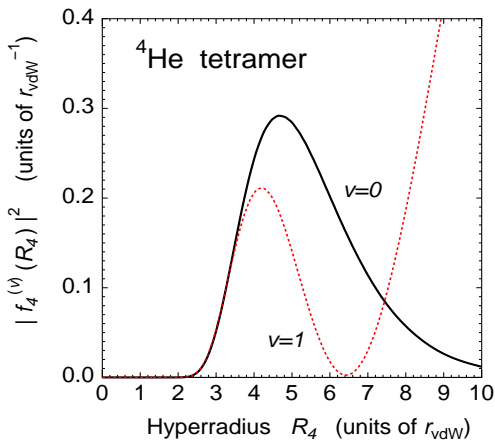,width=7.0cm,height=5.9cm}
\caption{(Color online) 
Probability density $|f_4^{(v)}(R_4)|^2 \:(v=0,1)$ of the $^4$He tetramer 
ground (solid curve) and excited  (dotted red curve) states
as a function of the  
hyperradius $R_4$ for \mbox{$R_4< 10 \,r_{\rm vdW}$.}
The density is normalized as $\int_0^\infty  |f_4^{(v)}(R_4)|^2 dR_4=1$.
The dotted curve has been multiplied by a factor 25 to emphasize that
the two tetramer states exhibit the same shape of 
the strong short-range correlations for $R_4 \lesssim 3.5 \, r_{\rm vdW}$.
}
\end{center}
\end{figure}

\begin{figure}[h]
\begin{center}
\epsfig{file=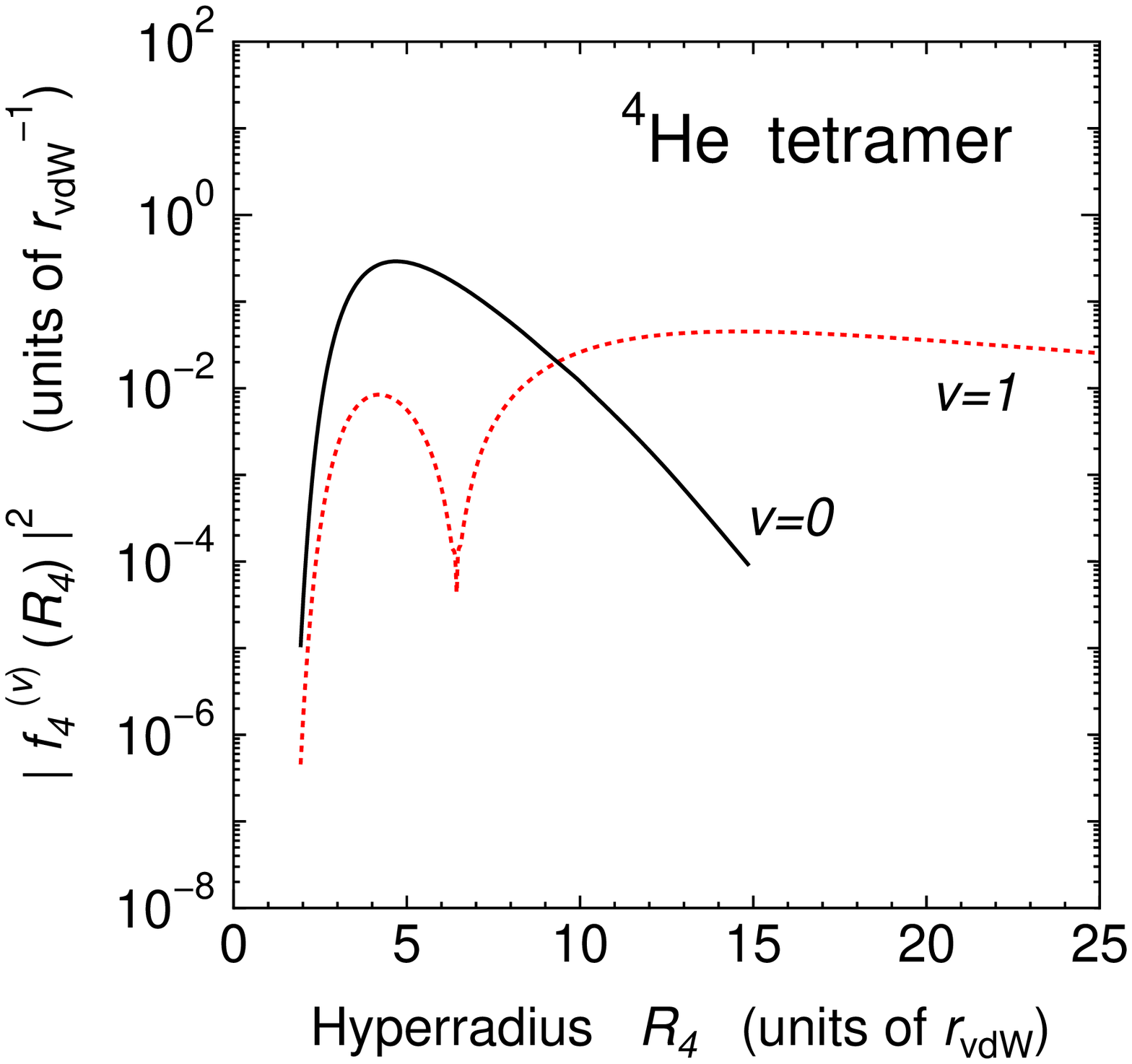,width=7.0cm,height=6.0cm}
\caption{(Color online) The same as Fig.~8
 for $R_4 < 25 \,r_{\rm vdW}$ in the log scale,
but no factor is multiplied to the dotted curve.
}
\end{center}
\end{figure}

\begin{figure}[h]
\begin{center}
\epsfig{file=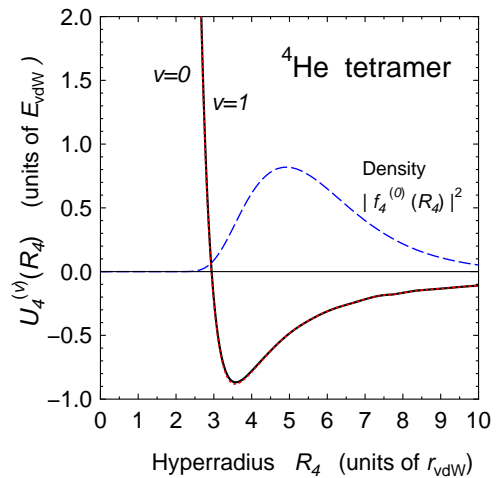,width=6.5cm,height=6.4cm}
\caption{(Color online) The four-body potential 
$U_4^{(v)}(R_4)\;(v=0,1)$ defined by
Eq.~(2.6) for the $^4$He tetramer ground (solid curve) and 
excited (dotted red curve) states.
The two curves overlap with each other within the thickness
of the lines. 
The dashed blue curve is for the density $|f_4^{(0)}(R_4)|^2$  
in arbitrary unit.
}
\end{center}
\end{figure}

The density distributions $|f_4^{(v)}(R_4)|^2$ are illustrated
in Fig.~8 for $R_4< 10 \,r_{\rm vdW}$ and 
in Fig.~9 (log scale) for \mbox{$R_4< 25 \, r_{\rm vdW}$} 
by the solid curve $(v=0)$ and the dotted red curve ($v=1$). 
The four-body potentials $U_4^{(v)}(R_4)$  
are shown in Fig.~10 as the solid curve $(v=0)$ and the
dotted red curve ($v=1$)~\footnote{   
Since the dotted curve becomes singular in the vicinity of
$R_4=6.4 \, r_{\rm vdW}$ where $f_4^{(1)}(R_4)=0$  
in Eq.~(2.6), we connected smoothly the lines on the both sides of the
vicinity. When the Schr\"{o}dinger equation~(2.5) is solved using this 
smoothed potential, the condition that the excited-state
wave function should have a node  at $R_4 =6.4 \, r_{\rm vdW}$ 
is guaranteed by the orthogonality to the ground-state wave function.
This situation is the same for the case of the trimer excited state
whose wave function $f_3^{(1)}(R_3)$  
has a node at $R_3=8.1 \, r_{\rm vdW}$ (Fig.~5).
}.    
We find that the condition $U_4^{(0)}(R_4) = U_4^{(1)}(R_4)$ is satisfied, 
which is required as the potential of the Schr\"{o}dinger 
equation (2.5) for the tetramer states attached to the trimer 
ground state; the two states are to be generated by the same hyperradial
potential dominantly for the relative motion between the trimer 
ground state and the fourth atom, for instance, as seen in Fig.~2 of 
Ref.~\cite{Stecher09}.

From the behavior of the potentials $U_4^{(v)}(R_4)$ and 
the density distributions $|f_4^{(v)}(R_4)|^2 \:(v=0,1)$, 
we recognize a strong repulsive barrier at $R_4 \approx 3\,r_{\rm vdW}$
that heavily suppresses the four-body density in the region 
$R_4 \lap 3\,r_{\rm vdW}$, which makes  
$a_-^{(4,0)}$ and $a_-^{(4,1)}$ insensitive to 
the short-range details of the interactions seen in Fig.~2.
The four-body repulsion radius  $R_4 \approx 3\,r_{\rm vdW}$
corresponds to the interparticle distance 
$r_{ij} \approx \sqrt{3} \,r_{\rm vdW}$ when all the $r_{ij}$ are
equal to each other in Eq.~(2.2).  The  distance
is $\sqrt{3}$ times larger than the two-body core radius 
$r_{12} \approx r_{\rm vdW}$ and is located outside the region
where the short-range details of the seven realistic potentials 
differ from each other (Fig.~2).
Inversely, if all the $r_{ij}$ 
are artificially replaced by the two-body core radius,
the four-body repulsion radius becomes 
$R_4 \approx \sqrt{3}\,r_{\rm vdW}$. The increase of the core radius
$R_4$ by factor $\sqrt{3}$ in the actual four-body system is    
due to the nonadiabatic four-body dynamics 
which is an extension of the mechanism 
discovered in Ref.~\cite{Wang12} for the three-body systems. 

For four-body bound states, the universal repulsion appears at 
larger distance than the three-body ones, suggesting that the four-body 
states tend to be more universal than the three-body ones. 
We note that in Ref.~\cite{Blume2000} for the study of $N$-body $^4$He 
clusters $(N=3-10)$ with the LM2M2 potential, 
effective $N$-body hyperradial potentials were approximately calculated 
on the basis of the Monte Carlo methods combined with the adiabatic 
hyperspherical approach and were used to calculate
the \mbox{$N$-body} excited bound states. 
The radius of the repulsive potential core becomes
larger with increasing $N$; the radius in the potential figure
seems consistent with the present result 
within $\sim$10\% for $N=3$ and $4$.
It is then conjectured that the universal $N$-body 
repulsion would generally appear in the Efimov associated 
$N$-body bound states $(N>3)$, rendering those states 
universal and insensitive to short-range details.

\section{Summary}

We have investigated the universality in  the $^4$He trimer and tetramer
using the seven realistic $^4$He potentials
and presented that even $^4$He potentials are consistent 
with measurements of broad Feshbach resonances in ultracold atoms, which 
shows the large extent of universality in three- and four-body systems 
with interactions featuring van der Waals tails.

We calculated the critical scattering lengths $a_-^{(v)}$ 
($a_-^{(4,v)}) \,(v=0,1)$ at which
the trimer (tetramer) energies cross 
the three-(four-)atom threshold.
From the nonadiabatic total wave function $\Psi_A^{(v)} (A=3,4)$ that is
described in terms of the full sets of the Jacobi coordinates,
we derived the hyperradial wave function $f_A^{(v)}(R_A)$
and  potential $U_A^{(v)}(R_A)$
as a function of the $A$-body hyperradius $R_A$.
The main conclusions are summarized as follows:

(i) We found the following universality in the $^4$He tetramer: 
The four-body hyperradial potentials $U_4^{(v)}(R_4)\:(v=0,1)$
have a repulsive barrier at the four-body hyperradius
\mbox{$R_4 \approx 3\, r_{\rm vdW}$,} 
which corresponds to the pair distance 
\mbox{$r_{ij} \approx \sqrt{3}\,r_{\rm vdW}$} 
when all the $r_{ij}$ are equal to each other. This pair distance is 
significantly larger than the radius \mbox{$r_{12} \approx r_{\rm vdW}$}
of the two-body repulsive core in the realistic $^4$He potentials.
Inside the barrier, \mbox{$R_4 \lap 3 \, r_{\rm vdW}$,}
the  probability density 
$|f_4^{(v)}(R_4)|^2$ to find the atoms is heavily suppressed.
The four-body barrier  prevents the particles  
from getting close together to explore non-universal features of 
the interactions at short distances; hence,
the critical scattering lengths are not affected by the difference 
in the short-range details of the interactions.
The seven realistic $^4$He potentials  
give the same value as 
\mbox{$a_-^{(4,0)}/\tilde{r}_{\rm vdW}=-4.69(1)$} 
and \mbox{$a_-^{(4,1)}/\tilde{r}_{\rm vdW}=-8.93(1)$}, 
which do not contradict 
the corresponding values obtained in the experiments in 
ultracold gases of the alkali-metal atoms. 

(ii) As for the universality in the $^4$He trimer,
we obtained the following result, which is consistent 
with that has been reported
in Refs.~\cite{Wang12,Endo12}:
The three-body hyperradial potentials  $U_3^{(v)}(R_3)\:(v=0,1)$
have a repulsive barrier at
$R_3 \approx 2 \, r_{\rm vdW}$, inside which 
the  probability density 
$|f_3^{(v)}(R_3)|^2$ of finding the atoms is heavily suppressed.
The hyperradius of the barrier corresponds to the pair distance 
$r_{ij} \approx \sqrt{2}\,r_{\rm vdW}$ when all the $r_{ij}$ are
equal to each other.  
The seven realistic $^4$He potentials  
give the same value of the universal three-body parameters
as \mbox{$a_-^{(0)}/\tilde{r}_{\rm vdW}=-9.78(1)$} and
\mbox{$a_-^{(1)}/\tilde{r}_{\rm vdW}=-165(1)$} 
independently of the short-range details of the  
potentials and are consistent with the corresponding values obtained 
in the ultracold-atom experiments.

(iii)  It will be one of the future subjects 
whether the four-body repulsive barrier at 
hyperradius $R_4 \approx 3 \, r_{\rm vdW}$ 
also appears for deeper pairwise interactions
that supports many two-body bound states like alkali atoms. 

\vspace{3mm}
\section*{Appendix}

We explain how to calculate  $\rho_3(R)$
of Eq.~(2.3) (the suffix 3 of $R_3$ is omitted).  
It is difficult to perform the integration  
(2.3) with the hyperradius $R$ treated explicitly since
$\Psi_3$ is not given as a function \mbox{of $R$}.
We expand $\rho_3(R)$ in terms of Gaussians
\begin{equation}
 \rho_3(R) = \sum_{n=1}^{n_{\rm max}} c_n e^{-\nu_n R^2} 
\end{equation}
\noindent
taking the range parameters \{$\nu_n$\} 
in a geometric progression (e.g. Eqs.~(2.10)--(2.13) in \mbox{paper I).}
The coefficients $c_n$ are obtained by solving a set of 
linear equations 
\begin{equation}
   \sum_{n=1}^{n_{\rm max}}  A_{in} c_n = B_i  \quad (i=1,...,n_{\rm max})
\end{equation}
\noindent
with the matrix elements
\begin{eqnarray}
A_{in}&=& \int_0^\infty e^{-(\nu_i + \nu_n)R^2} R^5 dR 
            = (\nu_i + \nu_n)^{-3} , \nonumber\\ 
B_i &= & \int_0^\infty  \rho_3(R')\, e^{-\nu_i R'^2} R'^5 dR'
  = \langle \,\Psi_3 \,| e^{- \nu_i R^2}  | \,\Psi_3\,\rangle
   \nonumber \\
&=&  \langle \,\Psi_3 \,| e^{- \frac{2}{3} \nu_i (r_{12}^2
+r_{23}^2+r_{31}^2)}  | \,\Psi_3\,\rangle .
\end{eqnarray}
$B_i$ is nothing but the expectation value of 
a three-body force of Gaussian shape, which is easily calculated.
We took the set of Gaussian ranges in a geometric progression
\{$n_{\rm max}=60, \,
R_1=0.2 \, r_{\rm vdW},\: R_{n_{\rm max}}=80 \, r_{\rm vdW}$\} where
$\nu_n=1/R_n^2$.  
This method also applies  to the $^4$He tetramer.

\section*{Acknowledgement}

The authors would like to thank \mbox{Dr.~S.~Endo for} valuable discussions
and careful reading the manuscript.
The numerical calculations were performed on \mbox{HITACHI SR16000}
at KEK and at YIFP in Kyoto University.


\end{document}